\documentclass{jfm}

\usepackage{graphicx}
\usepackage{newtxtext}
\usepackage{newtxmath}
\usepackage{natbib}
\usepackage{hyperref}
\usepackage{enumerate}
\usepackage{enumitem}
\usepackage[usenames, dvipsnames]{xcolor}
\hypersetup{
    colorlinks = true,
linkcolor=MidnightBlue,
    urlcolor   = blue,
    citecolor  = MidnightBlue,
}

\newcommand*{\F}{F}

\renewcommand*{\i}{\mathrm{i}}
\newcommand*{\im}{\mathrm{i}}
\newcommand*{\e}{\mathrm{e}}

\usepackage{esint} % For landdownint
\def\Xint#1{\mathchoice
   {\XXint\displaystyle\textstyle{#1}}%
   {\XXint\textstyle\scriptstyle{#1}}%
   {\XXint\scriptstyle\scriptscriptstyle{#1}}%
   {\XXint\scriptscriptstyle\scriptscriptstyle{#1}}%
   \!\int}
\def\XXint#1#2#3{{\setbox0=\hbox{$#1{#2#3}{\int}$}
     \vcenter{\hbox{$#2#3$}}\kern-.5\wd0}}
\def\dashint{\Xint-}

\newcommand{\RomanNumeralCaps}[1]
\linenumbers
\graphicspath{ {images/} }

\shorttitle{Exponential asymptotics and line vortices}
\shortauthor{Shelton and Trinh}

\title{Exponential asymptotics and the generation of free-surface flows by submerged line vortices}

\author{Josh Shelton\corresp{\email{j.shelton@bath.ac.uk}}
 \and Philippe H. Trinh\corresp{\email{p.trinh@bath.ac.uk}}}

\affiliation{
Department of Mathematical Sciences, University of Bath, Bath BA2 7AY, UK
}

\date{\today~[Draft]}

\begin{document}
\maketitle

\begin{abstract}
There has been significant recent interest in the study of water waves  coupled with non-zero vorticity. We derive analytical approximations for the exponentially-small free-surface waves generated in two-dimensions by one or several submerged point vortices when driven at low Froude numbers. The vortices are fixed in place, and a boundary-integral formulation in the arclength along the surface allows the study of nonlinear waves and strong point vortices. We demonstrate that for a single point vortex, techniques in exponential asymptotics prescribe the formation of waves in connection with the presence of Stokes lines originating from the vortex. When multiple point vortices are placed within the fluid, trapped waves may occur, which are confined to lie between the vortices. We also demonstrate that for the two-vortex problem, the phenomenon of trapped waves occurs for a countably infinite set of values of the Froude number. This work will form a basis for other asymptotic investigations of wave-structure interactions where vorticity plays a key role in the formation of surface waves.
\end{abstract}

\section{Introduction}\label{sec:intro}

In this paper we study the steady-state nonlinear flow of an ideal fluid past a submerged line vortex. As the vortices have fixed depth and horizontal displacement, they reduce to point vortices in the two-dimensional flows considered. The inviscid and incompressible fluid of infinite depth is assumed to be irrotational everywhere, with the exception of at the point vortices themselves. For a flow in the complex $z = x + \im y$-plane, with a vortex at $z = z_1$, the complex potential behaves as
\begin{equation}\label{eq:introf}
f = \phi + \im \psi \sim c z -\frac{\i \Gamma}{2 \pi} \log{(z-z_1)},    
\end{equation}
where $\Gamma$ is the circulation of the vortex, and the background flow is of speed $c$. The non-dimensional system is then characterised by two key parameters: $\Gamma_{\text{c}}=\Gamma/(cH)$, relating vortex strength(s) to inertial effects, and the Froude number, $F=c/\sqrt{g H}$, relating inertial effects to gravitational effects. Here, $H$ is the depth of the point vortex and $g$ is the constant acceleration due to gravity.

The study of such vortex-driven potential flows is complicated by the following fact.
The solution of two-dimensional ideal fluid-flow problems involves finding the velocity potential, $\phi$, and streamfunction, $\psi$, in terms of the coordinates $x$ and $y$, in the functional form of $f(z)$. However, it is often convenient to invert this dependency, instead calculating $z(f)$, so that the physical variables now have the forms $x(\phi,\psi)$ and $y(\phi,\psi)$. In this formulation, the free surface is a streamline along which $\psi$ is constant, so that the free surface is parameterised by $x(\phi)$ and $y(\phi)$. However, near the point vortex the local behaviour \eqref{eq:introf} can not be inverted analytically to give $z(f)$. This motivated the work of \cite{forbes1985effects}, who re-formulated the boundary-integral formulation in terms of a free-surface arclength, $s$, and a more complex set of governing equations results. 

\begin{figure}
\centering
\includegraphics[scale=1]{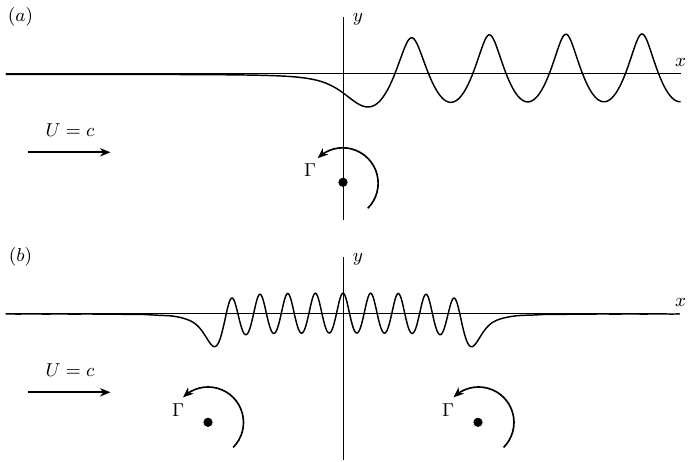}
\caption{\label{fig:C-physical} The two physical regimes of underlying point vortices considered within this paper are shown. In (a), a single point vortex with circulation $\Gamma$ is placed within the fluid. In (b), two point vortices, each with circulation $\Gamma$, are located at the same depth within the fluid. These solutions have been computed using the numerical scheme detailed in \S\ref{sec:results}.}
\end{figure}

\begin{figure}
\centering
\includegraphics[scale=1]{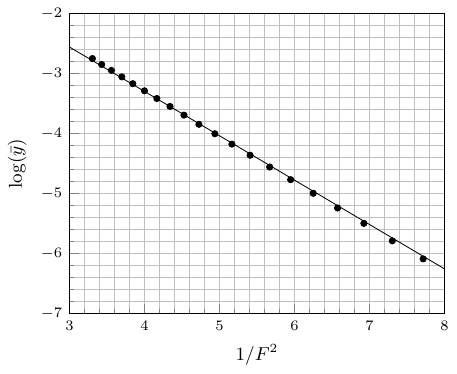}
\caption{\label{fig:A-expscale} The amplitude, $\bar{y}$, of the free-surface waves is shown for $\log(\bar{y})$ vs $1/F^2$ for the analytical (line) and numerical (dots) solutions of \S\ref{sec:asymptotics} and \S\ref{sec:results}. These have a fixed value of the nondimensional vortex strength, $\Gamma_{\text{c}}=0.25$. The graph confirms exponential smallness of the waves. The solid line has a gradient of $\approx 0.7395$, computed using the exponential asymptotic theory of \S\ref{sec:results}.}
\end{figure}

The imposition of a uniform stream as $x \to -\infty$ results in the generation of downstream free-surface waves, as shown in figure~\ref{fig:C-physical}(a). As hinted in the preliminary numerical investigations of \cite{forbes1985effects}, the wave amplitude tends to zero as $F \to 0$. In this work, we confirm this behaviour and demonstrate, both numerically and analytically, that the amplitude is exponentially-small in the low-Froude limit. For instance, the amplitude versus $1/F^2$ graph shown in figure~\ref{fig:A-expscale} demonstrates the fit between our asymptotic predictions of \S\ref{sec:asymptotics} and numerical results of \S\ref{sec:results}. We note that this theory is nonlinear in the vortex strength, $\Gamma_{\text{c}}$, and the assumption of small $\Gamma_{\text{c}}$ need not apply.

The purpose of this paper is to thus characterise the formation of water waves using the framework of exponential asymptotics. We show that these exponentially-small waves smoothly switch-on as the fluid passes beyond the vortex, resulting in oscillations as $x \to \infty$ in the far field. When two submerged vortices are considered, the waves switched-on due to each of the vortices may be out of phase with one another and cancel for certain values of the Froude number. This yields trapped waves between the vortices, and a free surface whose derivative decays to zero as $x \to \infty$. A trapped wave solution is depicted in figure~\ref{fig:C-physical}(b). This phenomenon of trapped waves has previously been studied for obstructions both within the fluid, and for flows of finite depth past lower topography. For instance, both \cite{sattar1973generation} and \cite{broeck1985waveless} detected these numerically for flows over a specified lower topography. More recent works, such as those by \cite{dias2004trapped}, \cite{hocking2013note}, and \cite{holmes2013waveless}, have focused on detecting parameter values for which these trapped wave solutions occur in various formulations.

The work in this paper provides a first step towards extending many of the existing ideas and techniques of exponential asymptotics, previously developed for purely gravity- or capillary-driven waves (e.g. \citealt{chapman_2002,chapman_2006}) to wave phenomena with vortices. As  noted above, because the governing equations require an alternative formulation (originally developed by \citealt{miksis1981axisymmetric}) the asymptotic formulation we present can be extended to other wave-structure interactions where the more general arc-length formulation of the water-wave equations is required. In addition, there has been significant recent interest in the study of water-wave phenomena with dominant vorticity effects, and we reference the recent extensive survey by \cite{haziot2022traveling} and references therein. The exponential-asymptotic techniques developed in this work can also be extended to situations where capillary ripples are forced on the surface of steep vortex-driven waves. The leading order solution in these asymptotic regimes would then be known analytically from the works of {\it e.g.} \citep{crowdy2010steady,crowdy2014hollow,crowdy2023exact}. These, and other exciting future directions, we shall discuss in \S\ref{sec:discussion}.

\section{Mathematical formulation and outline}\label{sec:formulation}
\noindent

\noindent We consider the typical configurations shown in \ref{fig:C-physical}. Following \cite{forbes1985effects}, in nondimensional form, the system is formulated in terms of the arclength, $s$, along the free surface, with unknown velocity potential $\phi = \phi(s)$, and free-surface positions, $(x(s), y(s))$. Then, the governing equations are given by Bernoulli's equation, an arclength relation between $x$ and $y$, and a boundary-integral equation. 

For a single submerged point vortex at $(x,y)=(0,-1)$, the three equations are
\begin{subequations}\label{eq:main}
\begin{align}
\label{eq:main1}
\frac{F^2}{2}\big[ \phi^{\prime}(s) \big]^2+y(s)=\frac{F^2}{2},\\
\label{eq:main2}
\big[x^{\prime}(s)\big]^2+\big[y^{\prime}(s)\big]^2=1,\\
\label{eq:main3}
\phi^{\prime}(s) x^{\prime}(s) -1 = \frac{\Gamma_{\text{c}}}{\pi}\frac{y(s)+1}{[x(s)]^2 + [y(s)+1]^2}+ \mathcal{I}[x,y,\phi^{\prime}].
\end{align}
\end{subequations}
In the above, two nondimensional parameters appear: the Froude number, $F$, and the vortex strength, $\Gamma_{\text{c}}$, defined by
\begin{equation}
\label{eq:constants}
   \F= \frac{c}{\sqrt{g H}} \quad \text{and} \quad \Gamma_{\text{c}}=\frac{\Gamma}{cH}.
\end{equation}
Here, $c$ is the speed of the fluid, $H$ is the depth of the submerged point vortex, $g$ is the constant acceleration due to gravity, and $\Gamma$ is the circulation of the point vortex. Furthermore, we have also introduced $\mathcal{I}$ as the nonlinear principle-valued integral defined by
\begin{equation}
\label{eq:Hint}
   \mathcal{I}[x,y,\phi^{\prime}]=\frac{1}{\pi} \dashint_{-\infty}^{\infty}\frac{[\phi^{\prime}(t)-x^{\prime}(t)][y(t)-y(s)]+y^{\prime}(t)[x(t)-x(s)]}{[x(t)-x(s)]^2+[y(t)-y(s)]^2} \, \mathrm{d} t.
\end{equation}

When the configuration with two point vortices is considered in \S\ref{sec:trappedexp}, the boundary-integral equation \eqref{eq:main3} will need to be modified to \eqref{eq:twovortex}.

\subsection{Analytic continuation}\label{sec:continuation}
In the exponential asymptotic procedure of \S\ref{sec:asymptotics}, we study the exponentially small terms that display the Stokes phenomenon across Stokes lines of the problem. These Stokes lines originate from singularities of the leading order asymptotic solution, which are located in the analytic continuation of the domain, the arclength $s$. The analytic continuation of the governing equations \eqref{eq:main1}-\eqref{eq:main3} is studied in this section.

We now analytically continue the domain $s \mapsto \sigma$, where $\sigma \in \mathbb{C}$. Bernoulli's equation \eqref{eq:main1} and the arclength relation \eqref{eq:main2} may be analytically continued in a straightforward manner, with all dependence on $s$ replaced by the complex valued variable $\sigma$. The analytic continuation of the boundary integral equation \eqref{eq:main3} is more complicated, due to the principal value integral $\mathcal{I}$ defined in \eqref{eq:Hint}. The analytic continuation of this integral is given by
\begin{equation}
\label{eq:Hintcomplex}
   \mathcal{I}[x,y,\phi^{\prime}]= \widehat{\mathcal{I}}[x,y,\phi^{\prime}]- a \i \phi^{\prime}(\sigma)y^{\prime}(\sigma),
\end{equation}
where $a=\pm 1$ denotes the direction of analytic continuation into $\text{Im}[\sigma]>0$ or $\text{Im}[\sigma]<0$, respectively, and $\widehat{\mathcal{I}}$ is the complex-valued integral. Equation \eqref{eq:Hintcomplex} may be verified by taking the limit of either $\text{Im}[\sigma] \to 0^+$, or $\text{Im}[\sigma] \to 0^-$, which yields half of a residue contribution associated with the singular point at $t=s$ of the integrand.

Substitution of \eqref{eq:Hintcomplex} into \eqref{eq:main3} then yields the analytically continued equations, given by
\begin{subequations}\label{eq:maincomplex}
\begin{align}
\label{eq:maincomplex1}
\frac{F^2}{2}\big[ \phi^{\prime}(\sigma) \big]^2+y(\sigma)=\frac{F^2}{2},\\
\label{eq:maincomplex2}
\big[x^{\prime}(\sigma)\big]^2+\big[y^{\prime}(\sigma)\big]^2=1,\\
\label{eq:maincomplex3}
\phi^{\prime}(\sigma) x^{\prime}(\sigma) -1 +a \i \phi^{\prime}(\sigma)y^{\prime}(\sigma) = \frac{\Gamma_{\text{c}}}{\pi}\frac{ y(\sigma)+1}{[x(\sigma)]^2 + [y(\sigma)+1]^2}+ \widehat{\mathcal{I}}[x,y,\phi^{\prime}].
\end{align}
\end{subequations}
The analytic continuation for situations with multiple point vortices is similarly done, with the only difference being the inclusion of additional point vortices in \eqref{eq:maincomplex3}.

\subsection{Outline of paper}\label{sec:outline}
In this work, we will consider the following two regimes depicted in figure~\ref{fig:C-physical}:
\begin{enumerate}[label=(\roman*),leftmargin=*, align = left, labelsep=\parindent, topsep=3pt, itemsep=2pt,itemindent=0pt ]
\item A single submerged point vortex, which is the formulation originally considered by \cite{forbes1985effects}. Imposing free stream conditions as $x \to -\infty$ results in surface waves generated by the vortex. Their amplitude is exponentially-small as $F \to 0$. This is the limit considered by \cite{chapman_2006} in the absence of vortical effects.
\item Two submerged point vortices of the same circulation. For certain critical values of the Froude number, $F$, the resultant waves are confined to lie between the two vortices. The amplitude of these is also exponentially small as $F \to 0$.
\end{enumerate}
We begin in \S\ref{sec:asymptotics} by determining these exponentially small waves using the techniques of exponential asymptotics. This relies on the optimal truncation of an algebraic asymptotic series for small Froude number, $F$, and deriving the connection of this to the Stokes phenomenon that acts on the exponentially small waves. The case for two submerged point vortices is then studied in \S\ref{sec:trappedexp}, where we derive the critical values of the Froude number for which the waves are trapped. Numerical solutions are computed in \S\ref{sec:results}, where comparison occurs with the exponential asymptotic predictions for the single vortex and double vortex cases.

\section{Exponential asymptotics}\label{sec:asymptotics}

\subsection{Early orders of the solution}\label{sec:lateterms}
We begin by considering the following asymptotic expansions, in powers of $F^2$, for the solutions, which are given by
\begin{equation}\label{eq:expansionsF}
x(\sigma)= \sum_{n=0}^{\infty} F^{2n}x_n(\sigma), \quad y(\sigma)= \sum_{n=0}^{\infty} F^{2n} y_n(\sigma), \quad \phi^{\prime}(\sigma)= \sum_{n=0}^{\infty}F^{2n}\phi_n^{\prime}(\sigma).
\end{equation}
Substitution of expansions \eqref{eq:expansionsF} into equations \eqref{eq:maincomplex1}-\eqref{eq:maincomplex3} yields at leading order three equations for the unknowns $x_0$, $y_0$, and $\phi_0^{\prime}$. The first of these, Bernoulli's equation \eqref{eq:maincomplex1}, yields $y_0(\sigma)=0$. This may be substituted into the second equation, \eqref{eq:maincomplex2}, to find $(x_0^{\prime})^2=1$, for which we consider $x_0^{\prime}=1$ without any loss of generality. This may be integrated to find $x_0=\sigma$, where the constant of integration has been chosen to set the origin at $x_0(0)=0$. Next, $\phi_0^{\prime}$ is determined from equation \eqref{eq:maincomplex3}. Since $y_0=0$, the integral $\widehat{\mathcal{I}}$ does not enter the leading order equation. This yields the leading order solutions as
\begin{equation}\label{eq:O1sols}
y_0(\sigma)=0, \qquad x_0(\sigma)=\sigma, \qquad \phi^{\prime}_0(\sigma) =1+\frac{\Gamma_{\text{c}}}{\pi}\frac{1}{(1+\sigma^2)}.
\end{equation}

Note that there is a singularity in $\phi_0^{\prime}$ above whenever $\sigma^2=-1$. This corresponds to the point vortex within the fluid at $\sigma=-\i$, as well as another singularity at $\sigma= \i$, which will produce a complex-conjugate contribution to the exponentially-small solution along the free surface.

Next at order $O(F^2)$, $y_1$ is found explicitly from \eqref{eq:maincomplex1}. We then find the equation $x_1^{\prime}=0$ from \eqref{eq:maincomplex2}, and $\phi_1$ is determined explicitly from \eqref{eq:maincomplex3}. This yields
\begin{equation}\label{eq:OFsols}
\left. \quad \begin{aligned}
y_1(\sigma)&=\frac{1}{2}\Big(1- \big(\phi_0^{\prime}\big)^2 \Big), \qquad x_1(\sigma)=0,\\
\phi^{\prime}_1(\sigma) &=-a \i \phi_0^{\prime}y_1^{\prime}+\frac{\Gamma_{\text{c}}(\sigma^2-1)}{\pi(1+\sigma^2)^2}y_1
+\widehat{\mathcal{I}}_1(\sigma),
\end{aligned} \quad \right\}
\end{equation}
where $\widehat{\mathcal{I}}_1$ is the $O(F^2)$ component of the complex-valued integral $\widehat{\mathcal{I}}$, originally defined along the real axis in equation \eqref{eq:Hint}.

\subsection{Late-term divergence}\label{sec:lateterms}
Our derivation of the exponentially-small terms and associated Stokes phenomenon of \S\ref{sec:smoothing} requires knowledge of the late-terms of the solution expansion \eqref{eq:expansionsF}, $x_n$, $y_n$, and $\phi_n^{\prime}$, as $n \to \infty$. We begin by determining the $O(F^{2n})$ components of equations \eqref{eq:maincomplex1}-\eqref{eq:maincomplex3}. The late-terms of Bernoulli's equation are given by
\begin{subequations} \label{eq:lateequations}
\begin{equation}\label{eq:lateequations1}
y_n + \phi_0^{\prime} \phi_{n-1}^{\prime} + \phi_1^{\prime}\phi_{n-2}^{\prime} +\cdots =0,
\end{equation}
for the arclength relation we have
\begin{equation}\label{eq:lateequations2}
x_0^{\prime}x_n^{\prime} + x_1^{\prime}x_{n-1}^{\prime}+\cdots + y_1^{\prime}y_{n-1}^{\prime}+ y_2^{\prime}y_{n-2}^{\prime}+ \cdots=0,\\
\end{equation}
and finally the boundary integral equation yields
\begin{multline}\label{eq:lateequations3}
x_0^{\prime}\phi_n^{\prime}+x_1^{\prime}\phi_{n-1}^{\prime}+\phi_0^{\prime}x_n^{\prime}+\cdots+a \i \big[\phi_0^{\prime}y_n^{\prime}+\phi_1^{\prime}y_{n-1}^{\prime}+y_1^{\prime}\phi_{n-1}^{\prime}+\cdots\big]\\
+\frac{\Gamma_{\text{c}}}{\pi}\bigg[\frac{y_n}{1+x_0^2}-\frac{2y_n}{(1+x_0^2)^2}+\cdots \bigg]-\widehat{\mathcal{I}}_n(\sigma)=0.
\end{multline}
\end{subequations}
In \eqref{eq:lateequations1}-\eqref{eq:lateequations3} above, only the terms that will appear at the first two orders of $n$ as $n \to \infty$ have been included. 

In \eqref{eq:lateequations3}, the $O(F^{2n})$ component of the complex-valued integral, $\widehat{\mathcal{I}}$ has been denoted by $\widehat{\mathcal{I}}_n$. The dominant components of this integral, as $n \to \infty$, require the integration of late-term asymptotic solutions that are either a function of the real valued integration domain, such as $y_n(t)$, or a function of the complex domain, such as $y_n(\sigma)$. The first of these, $y_n(t)$, is integrated along the real-valued free surface, away from any singular behaviour. It is thus subdominant to the other terms appearing in equation \eqref{eq:lateequations3}. This is analogous to neglecting the late terms of the complex-valued Hilbert transform in similar free-surface problems in exponential asymptotics [c.f. \cite{xie_2002}, \cite{chapman_2002}, \cite{chapman_2006}]. All that remains is to integrate the components of $\widehat{\mathcal{I}}_n$ that involve late-term solutions evaluated in the complex-valued domain. Of these, only that involving $y_n(\sigma)$ appears in the two leading orders, as $n \to \infty$, of equation \eqref{eq:lateequations3}. This component is given by
\begin{equation}\label{eq:OF2sols}
\widehat{\mathcal{I}}_n \sim -\frac{y_n(\sigma)}{\pi}\int_{-\infty}^{\infty} \frac{\phi_0^{\prime}(t)-1}{(t-\sigma)^2}\mathrm{d}t=-\frac{\Gamma_{\text{c}}}{\pi}\frac{y_n(\sigma)}{(\sigma+a \i)^2},
\end{equation}
for which the integral was evaluated by substituting for $\phi_0^{\prime}$ from equation \eqref{eq:O1sols}. Note that integration of $y_n(\sigma)$ was not required due to the lack of any dependence on the domain of integration, $t$.

Recall that the leading order solutions were singular at $\sigma = \pm i$. For each of the three solution expansions, this singularity first appeared in $\phi^{\prime}_0$, $y_1$, and $x_2$. Since successive terms in the asymptotic expansion involve differentiation of previous terms (for instance, equation \eqref{eq:lateequations1} for $y_n$ involves $\phi^{\prime}_{n-1}$, whose determination in equation \eqref{eq:lateequations3} requires knowledge of $y_{n-1}^{\prime}$), the strength of this singularity will grow as we proceed into the asymptotic series. Furthermore, this growing singular behaviour will also lead to the divergence of the late-term solutions as $n \to \infty$, which we capture analytically with the factorial-over-power ansatzes of
\begin{equation}\label{eq:ansatz}
x_n \sim X(\sigma)\frac{\Gamma(n+\alpha-1)}{[\chi(\sigma)]^{n+\alpha-1}}, \quad y_n \sim Y(\sigma)\frac{\Gamma(n+\alpha)}{[\chi(\sigma)]^{n+\alpha}}, \quad \phi_n \sim \Phi(\sigma)\frac{\Gamma(n+\alpha)}{[\chi(\sigma)]^{n+\alpha}}.
\end{equation}
Here, $\alpha$ is a constant, $\chi$ is the singulant function that will capture the singular behaviour of the solution at $\sigma=\pm \i$, and $X$, $Y$, and $\Phi$ are functional prefactors of the divergent solutions. It can be seen from the dominant balance as $n \to \infty$ of equations \eqref{eq:lateequations1} and \eqref{eq:lateequations2} that $x_{n+1} =O( y_{n})$ and $y_n=O( \phi_{n})$, which has motivated our precise ordering in $n$ in the ansatzes \eqref{eq:ansatz}.

Substitution of ansatzes \eqref{eq:ansatz} into the $O(F^{2n})$ equations \eqref{eq:lateequations1}-\eqref{eq:lateequations3} yields at leading order in $n$ the three equations
\begin{equation}\label{eq:chieq}
Y- \phi_0^{\prime}\chi^{\prime}\Phi=0, \qquad
\chi^{\prime}\Big(X+ y_1^{\prime} Y \Big)=0, \qquad
\chi^{\prime}\Big(\Phi+a \i \phi_0^{\prime}Y\Big)=0.
\end{equation}
While the last two of these equations permit the solution $\chi^{\prime}=0$, this is unable to satisfy the first equation in  \eqref{eq:chieq}. The remaining solutions can be solved to give $\chi^{\prime}= a \i (\phi_0^{\prime})^{-2}$, which we integrate to find
\begin{equation}\label{eq:chisol}
\chi_a(\sigma)=a \i  \int_{a \i}^{\sigma} \bigg[1+\frac{\Gamma_{\text{c}}}{\pi}\frac{1}{(1+t^2)}\bigg]^{-2}\mathrm{d} t.
\end{equation}
Here, we have introduced the notation $\chi_a=\chi$, where $a=\pm 1$, to discern between each singulant generated by the two singular points of $\phi_0^{\prime}$, which are given by $\sigma=\i$ and $\sigma= -\i$. The starting point of integration in \eqref{eq:chisol} is $\sigma = \pm \i$ to ensure that $\chi_a(a \i)=0$. This condition is required in order to match with an inner solution near this singular point. Integration of \eqref{eq:chisol} yields
\begin{equation}\label{eq:chisol1}
\begin{aligned}
\chi_a(\sigma)= &a \i \bigg[\sigma + \frac{\Gamma_{\text{c}}^2\sigma}{2(\Gamma_{\text{c}}+\pi)(\pi \sigma^2 + \Gamma_{\text{c}} + \pi)}-\frac{\Gamma_{\text{c}}  (3 \Gamma_{\text{c}}+4 \pi)}{2\sqrt{\pi}(\Gamma_{\text{c}}+\pi)^{3/2}}\tan^{-1}{\bigg(\frac{ \sqrt{\pi} \sigma}{\sqrt{ (\Gamma_{\text{c}}+\pi)}}\bigg)} \bigg]\\
&+1+\frac{\Gamma_{\text{c}}}{2(\Gamma_{\text{c}}+\pi)}-\frac{\Gamma_{\text{c}}(3 \Gamma_{\text{c}}+4 \pi)}{2 \sqrt{\pi}(\Gamma_{\text{c}}+\pi)^{3/2}} \tanh^{-1}{\bigg( \frac{\sqrt{\pi}}{\sqrt{\Gamma_{\text{c}}+\pi}}\bigg)}.
\end{aligned}
\end{equation}

\subsection{Solution of the late-term amplitude equations}\label{sec:ampsol}
We now determine the amplitude functions, $\Phi$, $X$, and $Y$, of the late term solutions. Note that if one of these amplitude functions is known, then the other two may be determined by the last two equations in \eqref{eq:chieq}. Thus, only one equation is required for the amplitude functions, which we find at the next order of $n$ in the late term equation \eqref{eq:lateequations1}.
This equation is given by
\begin{equation}\label{eq:ampeq1}
\phi_0^{\prime} \Phi^{\prime} = \phi_1^{\prime} \chi^{\prime} \Phi,
\end{equation}
which may be integrated to find the solution
\begin{equation}\label{eq:ampsol1}
\Phi(\sigma)=\Lambda \exp{\bigg( a \i \int_{0}^{\sigma} \frac{\phi_1^{\prime}(t)}{[\phi_0^{\prime}(t)]^3}\mathrm{d}t\bigg)}.
\end{equation}
In the above, $\Lambda$ is a constant of integration, which is determined by matching with an inner solution near the singular points $\sigma=a\i$. Once $\Phi$ is known, the remaining amplitude functions are determined by the equations $Y= a \i (\phi_0^{\prime})^{-1} \Phi$ and $X= a \i \phi_0^{\prime \prime} \Phi$.

We now calculate the constant, $\alpha$, that appears in the factorial-over-power ansatzes \eqref{eq:ansatz}.
This is determined by ensuring that the singular behaviour, as $\sigma \to a \i$, of each ansatz is consistent with the anticipated singular behaviours of
\begin{equation}\label{eq:expectedsingularity}
x_n = O\Big( (\sigma -a\i )^{1-3n} \Big), \qquad y_n = O\Big((\sigma- a \i)^{1-3n}\Big), \qquad \phi_n = O\Big( (\sigma -a \i)^{-3n} \Big).
\end{equation}
In taking the inner limit of $\Phi$ from \eqref{eq:ampsol1}, we have $\Phi = O(\sigma-a \i)^{3/2}$. Furthermore since $\chi = O\big((\sigma - a \i)^{3}\big)$, derived later in equation \eqref{eq:Apchi}, equating the power of the singularities for $\phi_n$ between the ansatz \eqref{eq:ansatz} and the anticipated singular behaviour above in \eqref{eq:expectedsingularity} yields the value of $\alpha=1/2$.
The constant of integration, $\Lambda$, that appears in solution \eqref{eq:ampsol1} for the amplitude function, $\Phi$, is derived in Appendix~\ref{sec:appinner} by matching the inner limit of the divergent solution, $\phi_n$, with an inner solution at $\sigma=a \i$. This yields
\begin{equation}\label{eq:constfop}
\alpha = \frac{1}{2} \qquad \text{and} \qquad \Lambda = -\frac{a \i \e^{-\mathcal{P}(a \i)}}{3 \sqrt{3}} \lim_{n \to \infty} \bigg(\frac{\hat{\phi}_n}{\Gamma(n+\alpha+1)}\bigg),
\end{equation}
where $\hat{\phi}_n$, determined via recurrence relation \eqref{eq:Apinnerrecrel}, is a constant appearing in the series expansion for the outer limit of the inner solution for $\phi$, and $\mathcal{P}(\sigma)$ is defined in equation \eqref{eq:strangeP}.

To conclude, the late-terms of the asymptotic expansions \eqref{eq:expansionsF} diverge in a factorial-over-power manner specified by the ansatzes \eqref{eq:ansatz}. Evaluation of this divergence requires the constants $\alpha$ and $\Lambda$ from equation \eqref{eq:constfop}, as well as the singulant function $\chi(\sigma)$ from \eqref{eq:chisol1} and amplitude function $\Phi(\sigma)$ from \eqref{eq:ampsol1}. These will be required in the derivation of the exponentially-small terms considered in the next section.

\subsection{Stokes smoothing and Stokes lines}\label{sec:smoothing}
The exponentially-small components of the solutions are now determined. We truncate the asymptotic expansions \eqref{eq:expansionsF} at $n=N-1$ and consider a remainder, yielding
\begin{equation}\label{eq:truncationsF}
x= \underbrace{\sum_{n=0}^{N-1} F^{2n}x_n}_{x_r}+~\bar{x}, \qquad y= \underbrace{\sum_{n=0}^{N-1} F^{2n} y_n}_{y_r} +~ \bar{y}, \qquad \phi^{\prime}= \underbrace{\sum_{n=0}^{N-1}F^{2n}\phi_n^{\prime}}_{\phi^{\prime}_r}+~\bar{\phi},
\end{equation}
where the truncated asymptotic expansions have been denoted by $x_r$, $y_r$, and $\phi_r^{\prime}$.
When $N$ is chosen optimally at the point at which the divergent expansions reorder as $n \to \infty$, given by
\begin{equation}\label{eq:optimalN}
N \sim \frac{\lvert \chi \rvert}{F^2} + \rho
\end{equation}
where $ 0\leq\rho <1$ to ensure that $N$ is an integer, the remainders to the asymptotic expansions \eqref{eq:truncationsF} will be exponentially-small.

Equations for these remainders are found by substituting the truncated expansions \eqref{eq:truncationsF} into the analytically continued equations \eqref{eq:maincomplex1}--\eqref{eq:maincomplex3}. These are given by
\begin{subequations}\label{eq:mainexp}
\begin{align}
\label{eq:mainexp1}
(F^2\phi_0^{\prime}+F^4\phi_1^{\prime})\bar{\phi}^{\prime}+\bar{y}&=-\xi_{\text{a}},\\
\label{eq:mainexp2}
2 \bar{x}^{\prime} +2F^2 y_1^{\prime}\bar{y}^{\prime} &=-\xi_{\text{b}},\\
\label{eq:mainexp3}
\bar{\phi}^{\prime}+a \i \phi_0^{\prime} \bar{y}^{\prime}&=-\xi_{\text{c}}.
\end{align}
\end{subequations}
In equations \eqref{eq:mainexp} above, nonlinear terms such as $\bar{x}^2$ were neglected as they will be exponentially subdominant. In anticipating that $\bar{x}=O(F^2\bar{y})=O(F^2\bar{\phi})$, terms of the first two orders of $F^2$ have been retained on the left-hand side of \eqref{eq:mainexp1}. Motivated by the late-term analysis, in which equations for the amplitude functions were obtained at leading order for the last two governing equations, we have only retained the leading order terms in equations \eqref{eq:mainexp2} and \eqref{eq:mainexp3}.
Furthermore, the forcing terms introduced in equations \eqref{eq:mainexp} are defined by
\begin{equation}\label{eq:mainforcing}
\left. \quad \begin{aligned}
\xi_{\text{a}}&=\frac{F^2}{2}(\phi_r^{\prime})^2+y_r-\frac{F^2}{2}, \qquad \xi_{\text{b}}=\big(x_r^{\prime}\big)^2+\big(y_r^{\prime}\big)^2-1,\\
\xi_{\text{c}}&=\phi_r^{\prime}x_r^{\prime}-1+a \i \phi_r^{\prime}y_r^{\prime}-\frac{\Gamma_{\text{c}}}{\pi}\frac{y_r+1}{(x_r)^2+(y_r+1)^2}-\widehat{\mathcal{I}}[x_r,y_r,\phi_r^{\prime}].
\end{aligned}\quad \right\}
\end{equation}
Since each order of these forcing terms will be identically zero up to and including $O(F^{2(N-1)})$, each of \eqref{eq:mainforcing} will be of $O(F^{2N})$. Only knowledge of $\xi_{\text{a}}$ will be required in the Stokes smoothing procedure of this section, and the leading-order component is given by
\begin{equation}
\label{eq:forcingscalings}
\xi_{\text{a}}\sim \phi_0^{\prime}\phi_{N-1}^{\prime}F^{2N}.
\end{equation}

Homogeneous solutions to equations \eqref{eq:mainexp}, for which the forcing terms on the right-hand sides are omitted, are
$\bar{x} \sim F^2 X\e^{-\chi/F^2}$, $\bar{y} \sim Y\e^{-\chi/F^2}$, and $\bar{\phi} \sim \Phi\e^{-\chi/F^2}$,
where the singulant $\chi$ and amplitude functions $X$, $Y$, and $\Phi$ satisfy the same equations as those found for the late-term solutions in \S\ref{sec:lateterms}. Next, we solve for the particular solutions of equations \eqref{eq:mainexp} through variation of parameters by multiplying the homogeneous solutions by an unknown function, $\mathcal{S}(\sigma)$, giving
\begin{equation}
\label{eq:varparam}
\left. \quad \begin{aligned}
 \bar{x} &\sim \mathcal{S}(\sigma)F^2X(\sigma)\e^{-\chi(\sigma)/F^2}, \\
 \bar{y} &\sim \mathcal{S}(\sigma)Y(\sigma)\e^{-\chi(\sigma)/F^2}, \\
 \bar{\phi} &\sim \mathcal{S}(\sigma)\Phi(\sigma)\e^{-\chi(\sigma)/F^2},
 \end{aligned}\quad \right\}
\end{equation}
where $Y=a \i \Phi / \phi_0^{\prime}$ and $X=-y_1^{\prime}Y$. The function $\mathcal{S}$ is called the Stokes multiplier as it will display the Stokes phenomenon across Stokes lines of the problem, which is demonstrated next. An equation for $\mathcal{S}$ is obtained by substituting \eqref{eq:varparam} into equation \eqref{eq:mainexp1}, yielding $F^2  \phi_0^{\prime}\Phi \e^{-\chi/F^2}\mathcal{S}^{\prime}(\sigma) \sim - \xi_{\text{a}}$. In substituting for the dominant behaviour of $\xi_{\text{a}}$ from \eqref{eq:forcingscalings} and the factorial-over-power divergence of $\phi^{\prime}_{N-1}$ from \eqref{eq:ansatz}, we change derivatives of $\mathcal{S}$ from $\sigma$ to $\chi$ to find
\begin{equation}
\label{eq:varparameq}
\frac{\mathrm{d}\mathcal{S}}{\mathrm{d} \chi} \sim \frac{\Gamma(N+\alpha)}{\chi^{N+\alpha}}F^{2(N-1)}\e^{\chi/F^2}.
\end{equation}
\begin{figure}
\centering
\includegraphics[scale=1]{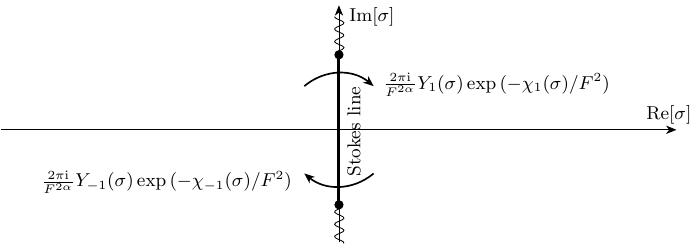}
\caption{\label{fig:E-Stokeslines} The Stokes lines (bold) lie along the imaginary axis between the two singular points of $\sigma=-\i$ and $\sigma=\i$. Branch cuts are shown with a wavy line.}
\end{figure}
In expanding as $N \to \infty$, and substituting for $N \sim \lvert \chi \rvert/F^2 + \rho$ from equation \eqref{eq:optimalN}, the right-hand side of equation \eqref{eq:varparam} is seen to be exponentially-small, except for in a boundary layer close to contours satisfying
\begin{equation}
\label{eq:dinglecond}
\text{Im}[\chi]=0 \quad \text{and} \quad \text{Re}[\chi]>0.
\end{equation}
These are the Stokes line conditions originally derived by \cite{dingle_book}. Across the Stokes lines, the solution for the Stokes multiplier $\mathcal{S}$,
\begin{equation}
\label{eq:varparamsol}
\mathcal{S}(\sigma)=S_a+\frac{\sqrt{2 \pi} \i}{F^{2 \alpha}} \int_{-\infty}^{\sqrt{\lvert \chi\rvert}\tfrac{\arg{(\chi)}}{F}} \exp{(-t^2/2)} \mathrm{d}t,
\end{equation}
rapidly varies from the constant $S_a$ to $S_a + 2 \pi \i / F^{2\alpha}$. This is the Stokes phenomenon, and the contours satisfying the Dingle conditions \eqref{eq:dinglecond} are shown in figure~\ref{fig:E-Stokeslines} to lie along the imaginary axis. For the one vortex case studied in this section, the upstream condition as $\text{Re}[\sigma] \to -\infty$ requires that $S_1=0$ and $S_{-1}=-2 \pi \i /F^{2 \alpha}$.

\subsection{Trapped waves generated by two submerged vortices}\label{sec:trappedexp}
We have so far studied the case of a single submerged point vortex. When multiple point vortices are placed within the fluid, the only change is to the boundary integral equation, previously specified in \eqref{eq:maincomplex3} for a single vortex. In this section we study the formulation of two submerged point vortices of the same nondimensional strength, $\Gamma_{\text{c}}$, located at $z=x+\i y = \pm\lambda - \i$, for which the analytically continued boundary integral equation is given by
\begin{equation}
\label{eq:twovortex}
 \begin{aligned}
\phi^{\prime}(\sigma) x^{\prime}(\sigma) -1 +a \i \phi^{\prime}(\sigma)y^{\prime}(\sigma) = \frac{\Gamma_{\text{c}}}{\pi}&\bigg[\frac{ y(\sigma)+1}{[x(\sigma)-\lambda]^2 + [y(\sigma)+1]^2}\\
&+ \frac{ y(\sigma)+1}{[x(\sigma)+\lambda]^2 + [y(\sigma)+1]^2}\bigg]+ \widehat{\mathcal{I}}[x,y,\phi].
 \end{aligned}
\end{equation}

Unlike the case for a single submerged point vortex that produces waves in the far field for $x \to \infty$, two identical point vortices can produce solutions for which the waves are confined to lie between the vortices, $-\lambda<\text{Re}[\sigma]<\lambda$. This occurs for critical values of the Froude number, which we now predict using the techniques of exponential asymptotics developed in the previous sections.

\begin{figure}
\centering
\includegraphics[scale=1]{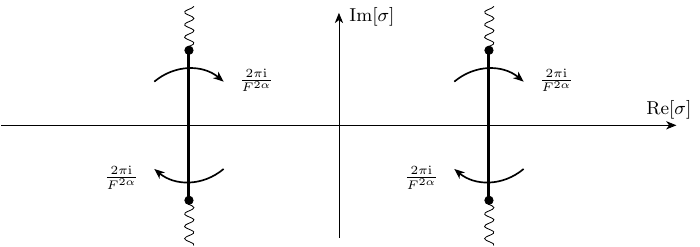}
\caption{\label{fig:F-Stokeslines} The Stokes lines (bold) generated by the four singular points are shown.}
\end{figure}

The first two orders of the asymptotic solution for $\phi$ are now given by
\begin{subequations}\label{eq:twovortexphi}
\begin{align}
\phi_0^{\prime}(\sigma)&=1+\frac{\Gamma_{\text{c}}}{\pi}\bigg[\frac{1}{1+(\sigma+\lambda)^2} +\frac{1}{1+(\sigma-\lambda)^2}\bigg],\\
\phi_1^{\prime}(\sigma)&=- a \i \phi_0^{\prime}y_1^{\prime}+\frac{\Gamma_{\text{c}}y_1}{\pi}\bigg[\frac{(\sigma+\lambda)^2-1}{[1+(\sigma+\lambda)^2]^2}+\frac{(\sigma-\lambda)^2-1}{[1+(\sigma-\lambda)^2]^2}\bigg]+\widehat{\mathcal{I}}_n(\sigma),
\end{align}
\end{subequations}
which are singular at the four locations $\sigma=-\lambda +a \i$ (from the vortex at $z=-\lambda-\i$) and $\sigma=\lambda + a \i$ (from the vortex at $z=\lambda -\i$). Note that we have again defined $a=\pm 1$ to indicate whether $\text{Im}[\sigma]>0$ or $\text{Im}[\sigma]<0$. These four singular points each have associated Stokes lines, shown in figure~\ref{fig:F-Stokeslines}.
In general, the waves switched on across the first Stokes lines, emanating from the points $\sigma=-\lambda+ a \i$, will be out of phase with the waves switched on across the second Stokes lines, from $\sigma=\lambda+a \i$. However, for certain values of $F$, the wave switched on across the first Stokes line is then switched off by the second Stokes line, yielding solutions with no waves for $\text{Re}[\sigma]> \lambda$. An example of this trapped solution was shown earlier in figure~\ref{fig:C-physical}(b).

Thus, in using the Stokes switching prediction for $\bar{\phi}$ shown in figure~\ref{fig:F-Stokeslines} and writing $\bar{y}= a \i \bar{\phi} / \phi_0^{\prime}$, we require for the two contributions of
\begin{equation}\label{eq:twovortexswitchings}
\left.
 \begin{aligned}
\bar{y}_1 &\sim -\frac{2 \pi}{F^{2 \alpha}\phi^{\prime}_0 }\Phi_1(\sigma) \exp{\Big(-\frac{\chi_1(\sigma)}{F^2}\Big)} + c.c., \\
\bar{y}_2 &\sim -\frac{2 \pi}{F^{2 \alpha}\phi^{\prime}_0 }\Phi_2(\sigma) \exp{\Big(-\frac{\chi_2(\sigma)}{F^2}\Big)} + c.c.,
 \end{aligned}
\quad  \right\}
\end{equation}
to cancel with one another for $\text{Re}[\sigma]>\lambda$. Here, we denoted $\chi_1$ and $\Phi_1$ as the singulant and amplitude function arising from the $\sigma=-\lambda+a \i$ singularities, and $\chi_2$ and $\Phi_2$ as those arising from the $\sigma=\lambda+a \i$ singularities.
The first of \eqref{eq:twovortexswitchings}, $\bar{y}_1$, is the contribution switched on as we pass from left to right across the Stokes lines associated with the singular points $\sigma=-\lambda+a\i$. The second, $\bar{y}_2$, is the contribution switched on from left to right by the Stokes lines associated with the $\sigma=\lambda+a\i$ singular point. Note that the specified contributions in \eqref{eq:twovortexswitchings} are from the $a=1$ contribution, and the unspecified complex-conjugate components are from that with $a=-1$.

We now simplify each of the expressions given in equation \eqref{eq:twovortexswitchings} by substituting for the amplitude functions $\Phi_1$ and $\Phi_2$, which satisfy the same equation as that found previously in \eqref{eq:ampeq1}. The only difference will be the constants of integration, which we denote by $\Lambda_1$ and $\Lambda_2$. This yields
\begin{equation}\label{eq:twovortexswitchingssimple}
\left.
 \begin{aligned}
\bar{\phi}_1 &\sim -\frac{4 \pi \lvert \Lambda_1 \rvert }{F^{2 \alpha}\phi_0^{\prime}}  \exp{\left(-\frac{\text{Re}[\chi_1]}{F^2}\right)}\cos{\left(\int_{0}^{\sigma} \frac{\phi_1^{\prime}(t)}{[\phi_0^{\prime}(t)]^3}\mathrm{d}t +\arg{[\Lambda_1]}-\frac{\text{Im}[\chi_1]}{F^2}\right)},\\
\bar{\phi}_2 &\sim -\frac{4 \pi \lvert \Lambda_2 \rvert }{F^{2 \alpha}\phi_0^{\prime}}  \exp{\left(-\frac{\text{Re}[\chi_2]}{F^2}\right)}\cos{\left(\int_{0}^{\sigma} \frac{\phi_1^{\prime}(t)}{[\phi_0^{\prime}(t)]^3}\mathrm{d}t +\arg{[\Lambda_2]}-\frac{\text{Im}[\chi_2]}{F^2}\right)}.
 \end{aligned}
\quad  \right\}
\end{equation}
Through integration of $\chi^{\prime}=a \i (\phi_0^{\prime})^{-2}$ and imposing the boundary conditions $\chi_1(a \i - \lambda)=0$ and $\chi_2(a \i + \lambda)=0$, it may be verified that along the free-surface, $\text{Im}[\sigma]=0$, we have  $\text{Re}[\chi_1]=\text{Re}[\chi_2]$. Furthermore, we also have $\lvert \Lambda_1 \rvert=\lvert \Lambda_2 \rvert$. This may be verified by matching with an inner solution, much like the procedure considered in Appendix~\ref{sec:appinner} for the case of a single point vortex. The same leading-order inner equation emerges regardless of the number of vortices considered, and so the outer limit of the inner solution is the same as in \eqref{eq:outerlimit} but with functional dependence on either $\sigma - \lambda -a \i$ or $\sigma + \lambda - a\i$. Thus, the only difference encountered in the matching procedure is in the inner limit of the outer divergent solution.

The prefactors multiplying each of the cosine functions in \eqref{eq:twovortexswitchingssimple} are identical, and the condition for them to cancel, $\bar{y}_1+\bar{y}_2=0$, yields
\begin{equation}
\label{eq:twovortexswitchings2}
\begin{aligned}
&\cos{\left( \int_0^{\sigma}\frac{\phi_1^{\prime}(t)}{[\phi_0^{\prime}(t)]^3}\mathrm{d}t + \frac{\arg{[\Lambda_1]} +\arg{[\Lambda_2]}}{2}-\frac{\text{Im}[\chi_1 + \chi_2]}{2F^2} \right)}\\
& \qquad \qquad \qquad \qquad \times \cos{\left(\frac{\arg{[\Lambda_1]} -\arg{[\Lambda_2]}}{2}-\frac{\text{Im}[\chi_1 - \chi_2]}{2F^2} \right)} =0.
\end{aligned}
\end{equation}
Note that since $\chi_1$ and $\chi_2$ satisfy the same differential equation, $\chi^{\prime}=a \i (\phi_0^{\prime})^{-2}$, originally derived in \S\ref{sec:lateterms}, the only difference between them are their constants of integration. Therefore $\text{Im}[\chi_1+\chi_2]$ will be a function of $\sigma$, and $\text{Im}[\chi_1-\chi_2]$ will be constant. Thus, only the second cosine component of \eqref{eq:twovortexswitchings2} is capable of satisfying the identity for $\text{Re}[\sigma]>\lambda$. Since this cosine function is zero when the argument equals $\pm \pi/2$, $\pm 3 \pi/2$, and so forth, we find
\begin{equation}
\label{eq:twovortexFconditions}
F_k = \sqrt{\frac{\text{Im}[\chi_1 - \chi_2]}{ \arg{[\Lambda_1]}-\arg{[\Lambda_2]}+\pi (2k+1)}}.
\end{equation}
for $k =0,1,2,\ldots$, and so forth.
Equation \eqref{eq:twovortexFconditions} yields the discrete values of the Froude number, $F_k$, for which the waves are confined to lie between the two submerged vortices. 

All that remains is to evaluate $\text{Im}[\chi_1-\chi_2]$, $\arg{[\Lambda_1]}$, and $\arg{[\Lambda_2]}$. Each of these singulants are found by integrating $\chi^{\prime}=a\i (\phi_0^{\prime})^{-2}$, where $\phi_0^{\prime}$ is specified in equation \eqref{eq:twovortexphi}, from the corresponding singular point. We may decompose each singulant into a real-valued integral along the Stokes line, and an imaginary-valued integral along the free-surface. Thus, $\text{Im}[\chi]$ is an integral along the free-surface, $\text{Im}[\sigma]=0$, from the intersection of the Stokes line to $\sigma$. This yields
\begin{equation}
\label{eq:twovortexchi}
\text{Im}[\chi_1(\sigma) - \chi_2(\sigma)] = \int_{-\lambda}^{\lambda}\bigg[1+\frac{\Gamma_{\text{c}}}{\pi}\bigg(\frac{1}{1+(t+\lambda)^2} +\frac{1}{1+(t-\lambda)^2}\bigg) \bigg]^{-2}\mathrm{d}t.
\end{equation}
In the numerical results of \S\ref{sec:numericaltrapped}, the integral in \eqref{eq:twovortexchi} is evaluated with a symbolic programming language.
Note that the Stokes lines depicted in figure~\ref{fig:F-Stokeslines} are not truly vertical, and are slightly curved such that they intersect the free surface at the points $-\lambda^*$ and $\lambda^*$. Thus, the range of integration in \eqref{eq:twovortexchi} should actually lie between $-\lambda^*<t<\lambda^*$; however since $\lambda^*$ is very close in value to $\lambda$ (for $\lambda=8$ and $\Gamma_c=0.3$, $\lambda^*\approx 7.99998$), this subtlety has been ignored. 

Comparisons between the analytical prediction of $F_k$ from \eqref{eq:twovortexFconditions} and numerical results are performed in \S\ref{sec:numericaltrapped}.

\section{Numerical results}\label{sec:results}
We begin in \S\ref{sec:numericalonevortexsols} by verifying with numerical results our analytical predictions for the exponentially-small scaling as $F \to 0$ for the case of a single vortex. This is given by the singulant function, $\chi$, from \eqref{eq:chisol1}, and comparisons are made for a range of values of the vorticity, $\Gamma_{\text{c}}$. The analytical predictions of the Froude numbers for trapped waves between two point vortices, given in \eqref{eq:twovortexFconditions}, are then compared to numerical predictions in \S\ref{sec:numericaltrapped}.

A detailed description of the numerical method used is given by \cite{forbes1985effects}, which we will briefly summarise here.
\begin{enumerate}[label=(\roman*),leftmargin=*, align = left, labelsep=\parindent, topsep=3pt, itemsep=2pt,itemindent=0pt ]
\item The real-valued domain, $s$, is truncated to lie between the values of $s_{\text{L}}$ and $s_{\text{R}}$. $N$ discretisation points are used, such that the numerical domain is given by $s_k=s_{\text{L}}+(k-1)(s_{\text{R}}-s_{\text{L}})/(N-1)$ for $1 \leq k \leq N$. The unknown solution is taken to be $y^{\prime}(s)$, which we define at each gridpoint by $y^{\prime}_k=y^{\prime}(s_k)$. The radiation conditions are imposed by enforcing $y_1=0$, $y_1^{\prime}=0$, $x_1^{\prime}=1$, $\phi_1^{\prime}=1$, $x_1=s_{\text{l}}$, and $\phi_1=s_{\text{L}}$, and the initial guess for $y_k^{\prime}$ is either zero or a previously computed solution.
\item Since we assume that $y^{\prime}_k$ is known at the next gridpoint, the arclength relation \eqref{eq:main2} yields $x^{\prime}_k$. Trapezoidal-rule integration then determines values for $x_k$ and $y_k$, which we use to find $\phi_k^{\prime}$ from Bernoulli's equation \eqref{eq:main1}. This process is repeated for $k=2$ to $k=N$ to find function values at every gridpoint. 
\item The boundary-integral equation \eqref{eq:main3} is evaluated at each gridpoint with the known values of $x_k$, $y_k$, $\phi^{\prime}_k$, $x^{\prime}_k$, and $y^{\prime}$. To avoid the singularity associated with the principal-valued integral $\mathcal{I}[x,y,\phi^{\prime}]$, each unknown that is not a function of the integration variable, $t$, is instead evaluated between gridpoints by interpolation.
\item This yields $N-1$ nonlinear equations from evaluating the boundary-integral equation between each gridpoint, $(s_k+s_{k+1})/2$, which is closed by the $N-1$ unknowns $y^{\prime}_k$ for $k=2$ to $k=N$. Solutions are found by minimising the residual through Newton iteration. For the trapped waves studied in \S\ref{sec:numericaltrapped}, we impose an additional constraint of symmetry about $s=0$ in the real-valued solution, $y(s)$, such that the Froude number, $F$, is determined as an eigenvalue.
\end{enumerate}

 \subsection{Waves generated by a single vortex}\label{sec:numericalonevortexsols}
 \begin{figure}
\centering
\includegraphics[scale=1]{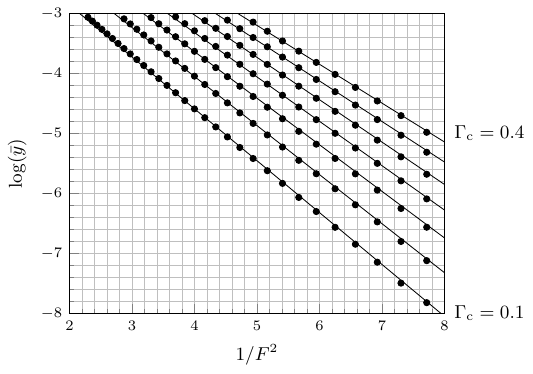}
\caption{\label{fig:B-expscale} The exponentially-small dependence of the wave amplitude is shown (dots) for numerical results for seven different values of $\Gamma_{\text{c}}=\{0.1,0.15,0.2,0.25,0.3,0.35,0.4\}$. Solid lines represent the analytical gradient found from the real part of $\chi$ in equation \eqref{eq:chisol1}. The behaviour of this gradient for different values of the vortex strength $\Gamma_{\text{c}}$ is shown in figure~\ref{fig:D-realchi}.}
\end{figure}

For the numerical results presented in this section, we have used $N=2000$ grid points, and a domain specified by $s_{\text{L}}=-40$ and $s_{\text{R}}=40$. In computing numerical solutions for a wide range of Froude numbers, and the values of $\Gamma_{\text{c}}=\{ 0.1, 0.15, 0.2, 0.25, 0.3, 0.35, 0.4\}$, the exponentially-small scaling as $F \to 0$ of the high-frequency waves present for $s>0$ may be measured. This is shown in the semilog plot of figure~\ref{fig:B-expscale}. We see that these lines, each of which represents solutions with a different value of $\Gamma_{\text{c}}$, are straight and thus the amplitude of these ripples is exponentially small as $F \to 0$. The gradient of each of these lines is expected to closely match the exponential scaling predicted analytically, given by the singulant $\chi$. Along the free surface, this is given by  $\text{Re}[\chi]$ from equation \eqref{eq:chisol1} which takes constant values.
\begin{figure}
\centering
\includegraphics[scale=1]{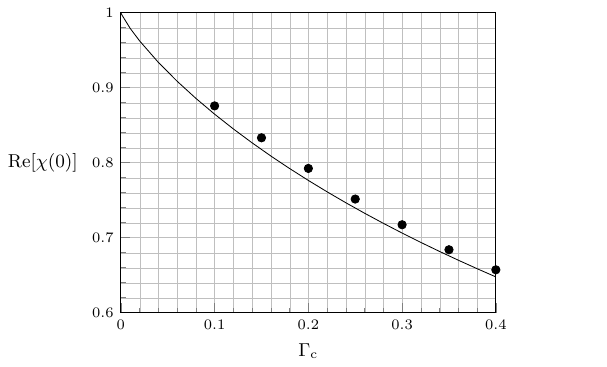}
\caption{\label{fig:D-realchi} The analytical prediction for $\text{Re}[\chi]$ along the free surface $\text{Im}[\sigma]=0$ from equation \eqref{eq:chisol1} is shown against the vorticity $\Gamma_{\text{c}}$ (line). The numerical predictions, corresponding to the slopes of the semilog plot in figure~\ref{fig:B-expscale}, are shown circled.}
\end{figure}
In figure~\ref{fig:D-realchi}, this analytical prediction is compared to the numerical values from figure~\ref{fig:B-expscale}, and good agreement is observed. Note that there are small instabilities present in the numerical solution which decay when the truncated domain is extended; upon which we expect the numerical results to tend towards the analytical prediction shown in figure~\ref{fig:D-realchi}.

 \begin{figure}
\centering
\includegraphics[scale=1]{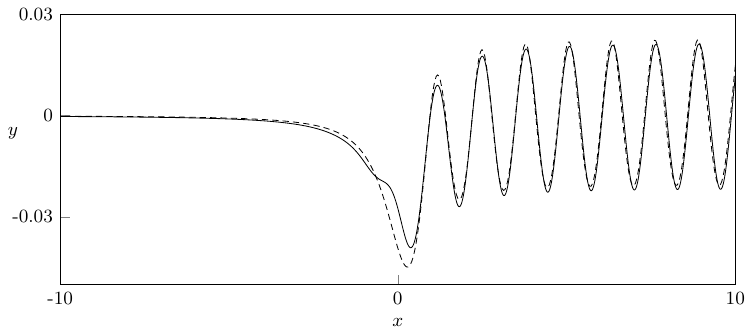}
\caption{\label{fig:H-onecomp} For $F=0.45$ and $\Gamma_{\text{c}}=0.4$, a numerical solution (dashed) is compared to an analytical solution (line) determined in \S\ref{sec:asymptotics}.}
\end{figure}
Comparison between a numerical and asymptotic solution profile is shown in figure~\ref{fig:H-onecomp} for $F=0.45$ and $\Gamma_{\text{c}}=0.4$. The numerical solution is determined by the scheme detailed at the beginning of \S\ref{sec:results}, with $N=2000$ discretisation points in the arclength, $-40 \leq s \leq 40$. The asymptotic solution plots $x(s)=x_0(s)+F^2x_1(s)+\bar{x}(s)$ against $y(s)=y_0(s) + F^2 y_1(s)+\bar{y}(s)$. These early order solutions, $x_0$, $x_1$, $y_0$, and $y_1$ are specified in equations \eqref{eq:O1sols} and \eqref{eq:OFsols}. The exponentially-small components, $\bar{x}$ and $\bar{y}$, are implemented from expression \eqref{eq:varparam}. This requires knowledge of the singulant, $\chi$, given in \eqref{eq:chisol1}, the amplitude functions $Y=a \i \Phi/\phi_0^{\prime}$ and $X=-y_1^{\prime}Y$ determined from $\Phi$ in \eqref{eq:ampsol1}, and the Stokes multiplier, $\mathcal{S}$, given in \eqref{eq:varparamsol}. A real-valued asymptotic solution is obtained through evaluating the sums $\bar{x}\rvert_{a=1}+\bar{x}\rvert_{a=-1}$ and $\bar{y}\rvert_{a=1}+\bar{y}\rvert_{a=-1}$ on the real-valued domain, $\sigma=s$, for $\text{Im}[\sigma]=0$. Note that in the determination of the constant $\Lambda$, its magnitude, $\lvert \Lambda \rvert$, has been fitted to equal that found from the corresponding numerical solution, and its argument (corresponding to a phase shift of the resultant wave) is determined from relation \eqref{eq:constfop} as $\text{arg}{[\Lambda]}=a\pi/2$.

 \subsection{Trapped gravity waves between two vortices}\label{sec:numericaltrapped}
 
\noindent We considered the case of two submerged point vortices analytically in \S\ref{sec:trappedexp}. When each vortex had the same nondimensional circulation, $\Gamma_{\text{c}}$, and depth equal to unity, trapped waves were seen to occur for certain discrete values of the Froude number, $F_k$. In this section, we compare the analytical prediction for $F_k$ from \eqref{eq:twovortexFconditions} with numerical results. These trapped numerical solutions are found with the method detailed at the beginning of \S\ref{sec:results}. In imposing the additional constraint of symmetry to eliminate waves downstream of the vortices, the special Froude number, $F_k$, is determined as an eigenvalue. These results were performed for $N=4000$ grid points, a domain between $s_{\text{L}}=-60$ and $s_{\text{R}}=60$, and horizontal vortex placement specified as $\lambda=8$.

\begin{figure}
\centering
\includegraphics[scale=1]{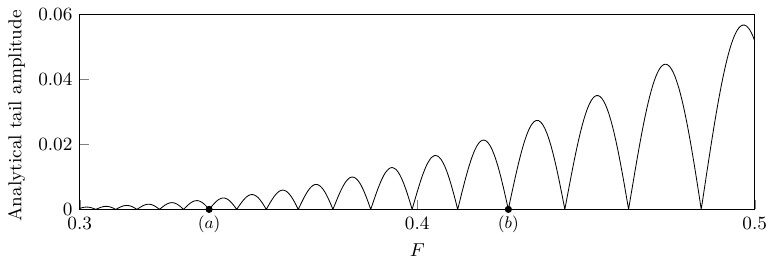}
\caption{\label{fig:I-twocompA} The amplitude of oscillations present for $s>\lambda$ in the asymptotic solutions is shown against the Froude number, $F$. Here, $\Gamma_{\text{c}}=0.3$ and $\lambda=8$. This amplitude is equal to zero at the locations $F_k$ derived in equation \eqref{eq:twovortexFconditions}. The two points marked (a) and (b) correspond to the profiles shown in figure~\ref{fig:I-twocompB}.}
\end{figure}
 \begin{figure}
\centering
\includegraphics[scale=1]{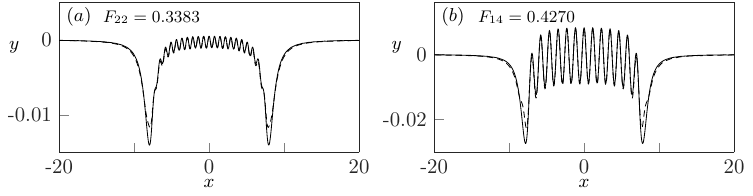}
\caption{\label{fig:I-twocompB} Two different trapped wave solutions are shown for $\Gamma_{\text{c}}=0.3$ and $\lambda=8$ corresponding to (a) $F = 0.3383$ and (b) $F = 0.4270$. Asymptotic solutions (solid line) are compared to numerical solutions (dashed) for (a) $k=22$ and (b) $k=14$. In each inset, the two curves are nearly indistinguishable to visual accuracy.}
\end{figure}
\begin{figure}
\centering
\includegraphics[scale=1]{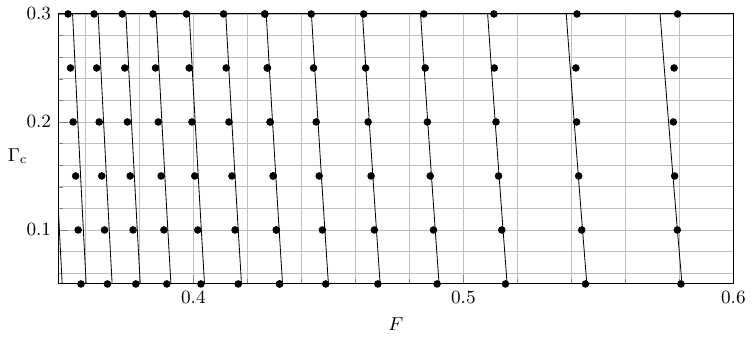}
\caption{\label{fig:G-FTrapped} Values of the Froude number, $F_k$, for which the waves are trapped between each submerged vortex are shown. The numerical results of \S\ref{sec:numericaltrapped} are shown by circles, and the analytical results from equation \eqref{eq:twovortexFconditions} are shown with lines. Here, $\lambda=8$, and for the numerical solutions $N=4000$, $s_{\text{L}}=-60$, and $s_{\text{R}}=60$.}
\end{figure}

In figure~\ref{fig:I-twocompA}, we plot the tail amplitude (for $s>\lambda$) of the asymptotic solutions for the values of $0.3<F<0.5$, $\Gamma_{\text{c}}=0.3$, and $\lambda=8$. This amplitude is equal to zero at the values of $F_k$ from equation \eqref{eq:twovortexFconditions}. The figure also contains additional markers denoted by (a), where $F = 0.3383$, and (b), where $F = 0.4270$. This corresponds to the figure~\ref{fig:I-twocompB} where we compare numerical solutions obtained in this section, and asymptotic solutions from \S\ref{sec:asymptotics} for those given values of $F$. The fit is excellent and the corresponding curves are nearly visually indistinguishable at the scale of the graphic.
Finally, in figure~\ref{fig:G-FTrapped}, we compare the values of $F_k$ obtained analytically and numerically. The straight lines are the analytical prediction from \eqref{eq:twovortexFconditions}, and dots represent the numerical values for $F_k$. The agreement between these is good for moderately small values of $F$; we notice that the error (horizontal distance) decreases as $F$ decreases from $0.6$ to around $0.4$. 

Numerical solutions of the double-vortex problem are particularly challenging, however, due to the finite-difference nature of the numerical scheme, and the truncation of the infinite domain. This seems to manifest in larger (relative) errors, particularly for $F \lessapprox 0.4$. We have verified that increasing the number of mesh points and increasing the domain size diminishes the numerical error. In addition, we have used the linear prediction of $F_k=\sqrt{\lambda /(k\pi+\pi/2)}$, to verify our analytical results for small values of $\Gamma_{\text{c}}$. Indeed, these numerical challenges are common in exponential asymptotic comparisons where it is necessary to use a sufficiently small value of the asymptotic parameter, but not-too-small so that numerical error overtakes the exponentially-small character of the solution.

\section{Conclusion}\label{sec:conclusion}
We have shown, through both numerical and analytical investigations, that the waves generated by submerged point vortices are exponentially small in the low-speed limit of $F \to 0$. Furthermore, when two submerged vortices are considered, oscillatory waves vanish downstream for certain values of the Froude number, $F$. Through the techniques of exponential asymptotics, we have demonstrated how these values may be derived. Their prediction relies on the understanding of singularities in the analytically continued domain that generate a divergent asymptotic expansion. The remainder to this series is exponentially small as $F \to 0$, and the study of the associated Stokes phenomenon yields discrete values of $F$ for which the waves are trapped between each vortex.

\section{Discussion}\label{sec:discussion}

The work presented here forms a basis for a number of interesting extensions involving exponentially-small water waves with gravity, capillarity, and/or vorticity providing singular perturbative effects. 

First, it should be remarked that the classical exponential-asymptotics theories by \emph{e.g.} \cite{chapman_2002,chapman_2006} for capillary- and gravity-driven surface waves produced in flows over topographies rely upon the existence of closed-form conformal maps. In such problems, the governing equations for the free-surface can be written in terms of a single complex-valued unknown (\emph{e.g.} the complex velocity), with the velocity potential serving as the independent variable. This includes situations such as flows past polygonal boundaries (related to the availability of the Schwarz-Christoffel mapping). The arclength formulation we have used in this work provides a more general setting for wave-structure interactions with arbitrary bodies, including for instance, flows past smoothed bodies specified in $(x,y)$-coordinates. Here, we have demonstrated that the exponential asymptotics can be generalised to such formulations. We expect that many of the interesting wave-structure interactions studied by \emph{e.g.} \cite{holmes2013waveless} (symmetric bottom topography), \cite{hocking2013note} (submerged semi-ellipse), and \cite{elcrat2006free} (submerged point vortex with lower topography) can be attacked using the technology we have developed here. 

Secondly, the phenomenon of trapped waves is an interesting one. The exponential asymptotics interpretation, whereby waves switched-on at one location (the Stokes line intersection) must be switched-off at another, provides an intuitive explanation for how trapped waves form in singularly perturbative limits. The context, in our problem, relates to vortices fixed within the fluid for modelling submerged obstructions, such as the submerged cylinders studied numerically by \cite{tuck1998tandem}. However, trapped waves have been detected numerically in other geometries including submerged bumps \citep{hocking2013note}, a semi-ellipse \citep{holmes2013waveless}, a trigonometric profile \citep{dias2004trapped}, spikes \citep{binder2005forced}, and a rectangular bump \citep{lustri2012free}. We expect that the `selection mechanism' that produces the countably infinite set of values \eqref{eq:twovortexFconditions} is a kind of universality in eigenvalue problems (cf. \citealt{chapman_2022_branch} for further discussion and examples). In our paper, we have mentioned some of the numerical challenges in verifying beyond-all-orders predictions, particularly in connection with finite-difference formulae and truncation of the infinite domain. In addition, we note that numerical verification may be further complicated by the fact that finite-difference discretisation introduces exponentially-small errors in the numerical solution, as shown by \cite{moston2023nanoptera}.

Finally, we note that in this paper, the forcing mechanism producing the waves was via the complex-plane singularities associated with the point vortices---then, we found that the waves were singularly perturbed due to the inertial term in Bernoulli's equation, thus producing exponentially small waves, scaling as $\exp(-\text{const.}/F^2)$. Recently, analytical solutions have been developed for pure-vorticity-driven water waves, notably in the works by \cite{crowdy2010steady,crowdy2014hollow,crowdy2023exact}. In essence, we believe these solutions can serve as leading-order approximations in the regime of small surface-tension; it might be expected that exponentially-small parasitic ripples then exist on the surface of such vorticity-driven profiles. This would then be similar to the work of \cite{shelton2021structure,shelton2022exponential} for parasitic capillary ripples on steep gravity waves. Numerical and analytical work on this class of problems is ongoing.

\mbox{}\par
{\bf \noindent Acknowledgements}.
We thank the anonymous referees for their helpful comments regarding the clarity of this work. 
We are grateful for many stimulating and motivating discussions that took place during the recent LMS-Bath symposium ``New Directions in Water Waves"  held at the University of Bath in July 2022. JS and PHT gratefully acknowledge support by the Engineering and Physical Sciences Research Council (EPSRC) [EP/V012479/1]. Finally, the authors would like to thank the Isaac Newton Institute for Mathematical Sciences, Cambridge, for support and hospitality during the programme Applicable Resurgent Asymptotics, where some work on this paper was undertaken (EPSRC grant no. [EP/R014604/1]). 

\mbox{}\par
{\bf \noindent Declaration of interests}. The authors report no conflict of interest.

\bibliographystyle{jfm}

\providecommand{\noopsort}[1]{}
\begin{thebibliography}{23}
\expandafter\ifx\csname natexlab\endcsname\relax\def\natexlab#1{#1}\fi
\def\au#1{#1} \def\ed#1{#1} \def\yr#1{#1}\def\at#1{#1}\def\jt#1{\textit{#1}}
  \def\bt#1{#1}\def\bvol#1{\textbf{#1}} \def\vol#1{#1} \def\pg#1{#1}
  \def\publ#1{#1}\def\arxiv#1{#1}\def\org#1{#1}\def\st#1{\textit{#1}}

\bibitem[Binder {\em et~al.\/}(2005)Binder, Vanden-Broeck \&
  Dias]{binder2005forced}
{\sc \au{Binder, B.~J.}, \au{Vanden-Broeck, J.-M.} \& \au{Dias, F.}} \yr{2005}
  \at{Forced solitary waves and fronts past submerged obstacles}.  \jt{Chaos}
  \bvol{15}~(3),  \pg{037106}.

\bibitem[Chapman {\em et~al.\/}(2022)Chapman, Dallaston, Kalliadasis, Trinh \&
  Witelski]{chapman_2022_branch}
{\sc \au{Chapman, S.~J.}, \au{Dallaston, M.~C.}, \au{Kalliadasis, S.},
  \au{Trinh, P.~H.} \& \au{Witelski, T.~P.}} \yr{2022}  \at{The role of
  exponential asymptotics and complex singularities in transitions and branch
  merging of nonlinear dynamics}.  \jt{In review.} .

\bibitem[Chapman \& Vanden-Broeck(2002)]{chapman_2002}
{\sc \au{Chapman, S.~J.} \& \au{Vanden-Broeck, J.-M.}} \yr{2002}
  \at{Exponential asymptotics and capillary waves}.  \jt{SIAM J. Appl. Math.}
  \bvol{62 (6)},  \pg{1872--1898}.

\bibitem[Chapman \& Vanden-Broeck(2006)]{chapman_2006}
{\sc \au{Chapman, S.~J.} \& \au{Vanden-Broeck, J.-M.}} \yr{2006}
  \at{Exponential asymptotics and gravity waves}.  \jt{J. Fluid Mech.}
  \bvol{567},  \pg{299--326}.

\bibitem[Crowdy(2023)]{crowdy2023exact}
{\sc \au{Crowdy, D.~G.}} \yr{2023}  \at{Exact solutions for steadily travelling
  water waves with submerged point vortices}.  \jt{J. Fluid Mech.}  \bvol{954},
   \pg{A47}.

\bibitem[Crowdy \& Nelson(2010)]{crowdy2010steady}
{\sc \au{Crowdy, D.~G.} \& \au{Nelson, R.}} \yr{2010}  \at{Steady interaction
  of a vortex street with a shear flow}.  \jt{Phys. Fluids}  \bvol{22}~(9),
  \pg{096601}.

\bibitem[Crowdy \& Roenby(2014)]{crowdy2014hollow}
{\sc \au{Crowdy, D.~G.} \& \au{Roenby, J.}} \yr{2014}  \at{Hollow vortices,
  capillary water waves and double quadrature domains}.  \jt{Fluid Dyn. Res.}
  \bvol{46}~(3),  \pg{031424}.

\bibitem[Dias \& Vanden-Broeck(2004)]{dias2004trapped}
{\sc \au{Dias, F.} \& \au{Vanden-Broeck, J.-M.}} \yr{2004}  \at{Trapped waves
  between submerged obstacles}.  \jt{J. Fluid Mech.}  \bvol{509},
  \pg{93--102}.

\bibitem[Dingle(1973)]{dingle_book}
{\sc \au{Dingle, R.~B.}} \yr{1973} {\em Asymptotic Expansions: Their Derivation
  and Interpretation\/}.  \publ{Academic Press, London}.

\bibitem[Elcrat \& Miller(2006)]{elcrat2006free}
{\sc \au{Elcrat, A.~R.} \& \au{Miller, K.~G.}} \yr{2006}  \at{Free surface
  waves in equilibrium with a vortex}.  \jt{Eur. J. Mech. B-Fluid.}
  \bvol{25}~(2),  \pg{255--266}.

\bibitem[Forbes(1985)]{forbes1985effects}
{\sc \au{Forbes, L.~K.}} \yr{1985}  \at{On the effects of non-linearity in
  free-surface flow about a submerged point vortex}.  \jt{J. Eng. Math.}
  \bvol{19}~(2),  \pg{139--155}.

\bibitem[Gazdar(1973)]{sattar1973generation}
{\sc \au{Gazdar, A.~S.}} \yr{1973}  \at{Generation of waves of small amplitude
  by an obstacle placed on the bottom of a running stream}.  \jt{J. Phys. Soc.
  Japan}  \bvol{34}~(2),  \pg{530--538}.

\bibitem[Haziot {\em et~al.\/}(2022)Haziot, Hur, Strauss, Toland, Wahl{\'e}n,
  Walsh \& Wheeler]{haziot2022traveling}
{\sc \au{Haziot, S.}, \au{Hur, V.}, \au{Strauss, W.}, \au{Toland, J.},
  \au{Wahl{\'e}n, E.}, \au{Walsh, S.} \& \au{Wheeler, M.}} \yr{2022}
  \at{Traveling water waves—the ebb and flow of two centuries}.  \jt{Quart.
  Appl. Math.}  \bvol{80}~(2),  \pg{317--401}.

\bibitem[Hocking {\em et~al.\/}(2013)Hocking, Holmes \&
  Forbes]{hocking2013note}
{\sc \au{Hocking, G.~C.}, \au{Holmes, R.~J.} \& \au{Forbes, L.~K.}} \yr{2013}
  \at{A note on waveless subcritical flow past a submerged semi-ellipse}.
  \jt{J. Eng. Math.}  \bvol{81}~(1),  \pg{1--8}.

\bibitem[Holmes {\em et~al.\/}(2013)Holmes, Hocking, Forbes \&
  Baillard]{holmes2013waveless}
{\sc \au{Holmes, R.~J.}, \au{Hocking, G.~C.}, \au{Forbes, L.~K.} \&
  \au{Baillard, N.~Y.}} \yr{2013}  \at{Waveless subcritical flow past symmetric
  bottom topography}.  \jt{Eur. J. Appl. Math.}  \bvol{24}~(2),  \pg{213--230}.

\bibitem[Lustri {\em et~al.\/}(2012)Lustri, McCue \& Binder]{lustri2012free}
{\sc \au{Lustri, C.~J.}, \au{McCue, S.~W.} \& \au{Binder, B.~J.}} \yr{2012}
  \at{Free surface flow past topography: a beyond-all-orders approach}.
  \jt{Euro. J. Appl. Math.}  \bvol{23}~(4),  \pg{441--467}.

\bibitem[Miksis {\em et~al.\/}(1981)Miksis, Vanden-Broeck \&
  Keller]{miksis1981axisymmetric}
{\sc \au{Miksis, M.}, \au{Vanden-Broeck, J.-M.} \& \au{Keller, J.~B.}}
  \yr{1981}  \at{Axisymmetric bubble or drop in a uniform flow}.  \jt{J. Fluid
  Mech.}  \bvol{108},  \pg{89--100}.

\bibitem[Moston-Duggan {\em et~al.\/}(2023)Moston-Duggan, Porter \&
  Lustri]{moston2023nanoptera}
{\sc \au{Moston-Duggan, A.~J.}, \au{Porter, M.~A.} \& \au{Lustri, C.~J.}}
  \yr{2023}  \at{Nanoptera in higher-order nonlinear schr{\"o}dinger equations:
  Effects of discretization}.  \jt{J. Nonlinear Sci.}  \bvol{33}~(1),
  \pg{1--47}.

\bibitem[Shelton {\em et~al.\/}(2021)Shelton, Milewski \&
  Trinh]{shelton2021structure}
{\sc \au{Shelton, J.}, \au{Milewski, P.} \& \au{Trinh, P.~H.}} \yr{2021}
  \at{On the structure of steady parasitic gravity-capillary waves in the small
  surface tension limit}.  \jt{J. Fluid Mech.}  \bvol{922}.

\bibitem[Shelton \& Trinh(2022)]{shelton2022exponential}
{\sc \au{Shelton, J.} \& \au{Trinh, P.H.}} \yr{2022}  \at{Exponential
  asymptotics for steady parasitic capillary ripples on steep gravity waves}.
  \jt{J. Fluid Mech.}  \bvol{939}.

\bibitem[Tuck \& Scullen(1998)]{tuck1998tandem}
{\sc \au{Tuck, E.~O.} \& \au{Scullen, D.~C.}} \yr{1998}  \at{Tandem submerged
  cylinders each subject to zero drag}.  \jt{J. Fluid Mech.}  \bvol{364},
  \pg{211--220}.

\bibitem[Vanden-Broeck \& Tuck(1985)]{broeck1985waveless}
{\sc \au{Vanden-Broeck, J.-M.} \& \au{Tuck, E.~O.}} \yr{1985}  \at{Waveless
  free-surface pressure distributions}.  \jt{J. Ship Res.}  \bvol{29}~(03),
  \pg{151--158}.

\bibitem[Xie \& Tanveer(2002)]{xie_2002}
{\sc \au{Xie, X.} \& \au{Tanveer, S.}} \yr{2002}  \at{Analyticity and
  nonexistence of classical steady {H}ele-{S}haw fingers}.  \jt{Commun. Pur.
  Appl. Math.}  \bvol{56}~(3),  \pg{353--402}.

\end{thebibliography}
\providecommand{\noopsort}[1]{}

\appendix
\section{Inner analysis at the singularities $\sigma=\pm \i$}\label{sec:appinner}

In order to determine the constant of integration of the amplitude function $\Phi(\sigma)$ from equation \eqref{eq:ampeq1}, knowledge of the inner solutions at the singularities $\sigma= \i$ and $\sigma=-\i$ is required. In this section, we study the inner boundary layer at both of these locations, for which matching with the inner limit of the outer solutions determines the constant of integration.

First, we note that in the outer region, where $\sigma=O(1)$, the asymptotic series first reorder whenever
\begin{equation}\label{eq:Apreordering}
\phi_0^{\prime}(\sigma) \sim F^2 \phi_1^{\prime}(\sigma), \qquad y_1(\sigma) \sim F^2 y_2(\sigma), \qquad x_2(\sigma) \sim F^2 x_3(\sigma).
\end{equation}
In substituting for the early orders of the asymptotic solutions specified in equations \eqref{eq:O1sols}, \eqref{eq:OFsols}, and \eqref{eq:OF2sols}, we see that each of \eqref{eq:Apreordering} reorder in a boundary layer of the same width, given by $\sigma-a \i =O(F^{2/3})$. We thus introduce the inner variable, $\hat{\sigma}$, by the relation
\begin{equation}\label{eq:Apinnervar}
\sigma-a \i=\hat{\sigma}F^{2/3},
\end{equation}
for which $\hat{\sigma}=O(1)$ in the inner region. Since the asymptotic series each reorder near the two locations of $\sigma=\i$ and $\sigma=-\i$, we have again used the notation $a=\pm 1$ to distinguish between these two cases.

Next, to determine the form of the inner solutions, we take the inner limit of the outer series expansions for $\phi^{\prime}$, $x$, and $y$, by substituting for the inner variable $\hat{\sigma}$ defined in \eqref{eq:Apinnervar} and expanding as $F \to 0$. This yields
\begin{equation}\label{eq:Apinnerlimits}
\phi^{\prime} \sim \frac{1}{F^{2/3}} \bigg[-\frac{a \i \Gamma_{\text{c}}}{2 \pi}\frac{1}{\hat{\sigma}}  +\cdots\bigg], \quad y \sim F^{2/3} \bigg[ \frac{\Gamma_{\text{c}}^2}{8 \pi^2}\frac{1}{\hat{\sigma}^2}+\cdots\bigg], \quad x \sim a \i + F^{2/3} \bigg[\hat{\sigma}+\cdots \bigg],
\end{equation}
where the omitted terms, represented by ($\cdots$), are from the inner limit of lower order terms of the outer asymptotic expansion. For instance, the next term in the inner limit of $\phi^{\prime}$ is of $O(F^{-2/3}\hat{\sigma}^{-4})$. The form of the inner limits in \eqref{eq:Apinnerlimits} motivates our definition of the inner solutions, $\hat{\phi}(\hat{\sigma})$, $\hat{y}(\hat{\sigma})$, and $\hat{x}(\hat{\sigma})$, through the equations
\begin{equation}\label{eq:Apinnersols}
\phi^{\prime}=-\frac{a \i \Gamma_{\text{c}}}{2 \pi F^{2/3}} \frac{\hat{\phi}(\hat{\sigma})}{\hat{\sigma}},  \qquad  y=\frac{\Gamma_{\text{c}}^2F^{2/3}}{8 \pi^2}\frac{\hat{y}(\hat{\sigma})}{\hat{\sigma}^2},\qquad x= a \i + \hat{\sigma}F^{2/3}\hat{x}(\hat{\sigma}).
\end{equation}
The form of the inner variables introduced in \eqref{eq:Apinnersols} ensures that the first term in the series expansion for their outer limit will be equal to unity. Furthermore, based on the form of the inner limit of the singulant, $\chi$, from equation \eqref{eq:chisol},
\begin{equation}\label{eq:Apchi}
\chi \sim -\frac{4 a \i \pi^2}{3\Gamma_{\text{c}}^2} \hat{\sigma}^3 F^2,
\end{equation}
the outer limit of the inner solutions will be a series expansion in inverse powers of $-4 a \i \pi^2 \hat{\sigma}^3/(3 \Gamma_{\text{c}}^2)$. We thus introduce the variable $z$, defined by
\begin{equation}\label{eq:Apnewinnervar}
z=-\frac{4 a \i \pi^2}{3\Gamma^2_{\text{c}}} \hat{\sigma}^3,
\end{equation}
to ensure that these series expansions are in inverse powers of $z$ alone.

\subsection{Inner equation}\label{sec:Apinnereq}
The leading order inner equations, as $F \to 0$, may now be derived by substituting \eqref{eq:Apinnersols} for the inner variables into the outer equations \eqref{eq:maincomplex1}-\eqref{eq:maincomplex3}, yielding
\begin{subequations}\label{eq:maininner}
\begin{align}
\label{eq:maininner1}
\hat{y}-\hat{\phi}^2=0,\\
\label{eq:maininner2}
\Big(\hat{x} + 3 z \hat{x}^{\prime}\Big)^2-\bigg(\frac{1}{3z}\hat{y}-\frac{1}{2}\hat{y}^{\prime} \bigg)^2=1,\\
\label{eq:maininner3}
\hat{\phi}\bigg(\hat{x} -\frac{1}{6 z}\hat{y}\bigg)\bigg(\hat{x}+3 z\hat{x}^{\prime}-\frac{1}{3z}\hat{y}+\frac{1}{2}\hat{y}^{\prime} \bigg)=1.
\end{align}
\end{subequations}
The inner solutions, $\hat{\phi}(z)$, $\hat{y}(z)$, and $\hat{x}(z)$, will satisfy equations \eqref{eq:maininner1}-\eqref{eq:maininner3}. Rather than solve these inner equations exactly, knowledge of the inner solutions is only required under the outer limit of $z \to \infty$ in order to match with the inner limit of the outer solutions to determine their divergent form. Thus, we will consider the following series expansions for these inner unknowns,
\begin{equation}\label{eq:Apinnerexpansions}
\hat{\phi}(z)= \sum_{n=0}^{\infty} \frac{\hat{\phi}_n}{z^n}, \qquad \hat{y}(z)= \sum_{n=0}^{\infty}\frac{\hat{y}_n}{z^n}, \qquad \hat{x}(z)= \sum_{n=0}^{\infty} \frac{\hat{x}_n}{z^n},
\end{equation}
which hold as $z \to \infty$.

At leading order as $z \to \infty$ we have, by the definition on the inner solutions in equation \eqref{eq:Apinnersols},
\begin{equation}\label{eq:Apinnerexpansionsfirst}
\hat{\phi}_0 =1, \qquad \hat{y}_0 =1, \qquad \hat{x}_0 =1.
\end{equation}
Determination of $\hat{\phi}_n$, $\hat{y}_n$, and $\hat{x}_n$, as $n \to \infty$, requires the evaluation of a recurrence relation, which is now given. Firstly, substitution of expansions \eqref{eq:Apinnerexpansions} into the inner equation \eqref{eq:maininner2} yields
\begin{subequations}\label{eq:Apinnerrecrel}
\begin{equation}\label{eq:Apinnerrecrel2}
\begin{aligned}
\hat{x}_1&=0, \\
\hat{x}_n &= \sum_{m=1}^{n-1}\frac{(3m-1)(3n-3m-1)}{2(1-3n)} \left(\frac{\hat{y}_{m-1}\hat{y}_{n-m-1}}{36}-\hat{x}_{m}\hat{x}_{n-m} \right) \qquad \text{for $~ n \geq 2$}.
\end{aligned}
\end{equation}
Next, we substitute the same expansions into the inner equation \eqref{eq:maininner3}, yielding
\begin{equation}\label{eq:Apinnerrecrel3}
\begin{aligned}
\hat{\phi}_1=\frac{1}{2}, \qquad \hat{\phi}_n &=\frac{1}{36}\sum_{m=2}^{n} \sum_{q=1}^{m-1}(3q-1)\left(6\hat{x}_q +\hat{y}_{q-1} \right) \left( 6\hat{x}_{m-q}-\hat{y}_{m-q-1}\right)\hat{\phi}_{n-m}\\
& ~~~ -\frac{1}{2}\sum_{m=1}^{n} \Big((4-6m)\hat{x}_m -m\hat{y}_{m-1}\Big)\hat{\phi}_{n-m} \qquad \text{for $~ n \geq 2$}.
\end{aligned}
\end{equation}
Lastly, a recurrence relation for $\hat{y}_n$ is found from equation \eqref{eq:maininner1} to be
\begin{equation}\label{eq:Apinnerrecrel1}
\hat{y}_1=1, \qquad \hat{y}_n = \sum_{m=0}^{n}\hat{\phi}_m \hat{\phi}_{n-m} \qquad \text{for  $~n \geq 2$}.
\end{equation}
\end{subequations}
Assuming that $\hat{\phi}_{n-1}$, $\hat{y}_{n-1}$, and $\hat{x}_{n-1}$ are known, $\hat{x}_n$ can be determined from equation \eqref{eq:Apinnerrecrel2}, which then yields a value for $\hat{\phi}_n$ from equation \eqref{eq:Apinnerrecrel3}. Lastly, $\hat{y}_n$ is found by evaluating equation \eqref{eq:Apinnerrecrel1}.

\subsection{Matching and determination of the constant $\Lambda$}
We now match the outer limit of the inner solution, $\hat{\phi}$, with the inner limit of the outer solution, $\phi^{\prime}$. In writing the outer limit of the inner solution in outer variables, we have
\begin{equation}\label{eq:outerlimit}
\phi^{\prime}= \frac{-a \i \Gamma_{\text{c}}}{2 \pi}\sum_{n=0}^{\infty} \frac{F^{2n}\hat{\phi}_n}{\Big(-\frac{4 a \i \pi^2}{3 \Gamma_{\text{c}}^2}\Big)^n(\sigma-a \i)^{3n+1}},
\end{equation}
and for the inner limit of the outer solution,
\begin{equation}\label{eq:innerlimit}
\begin{aligned}
\phi^{\prime}= \sum_{n=0}^{\infty} F^{2n} \phi_n^{\prime} &\sim \sum_{n=0}^{\infty} -F^{2n} \chi^{\prime} \Phi \frac{\Gamma(n+\alpha+1)}{\chi^{n+\alpha+1}}\\
& \sim \sum_{n=0}^{\infty}-\frac{4 \pi^2 \Lambda}{\Gamma_{\text{c}}^2(-a\i)^{1/2}} \e^{\mathcal{P}(a \i)}\frac{F^{2n}\Gamma{(n+\alpha+1)}}{\Big(-\frac{4 a \i \pi^2}{3 \Gamma_{\text{c}}^2}\Big)^{n+\alpha+1}(\sigma-a \i)^{3n+3\alpha-1/2}}.
\end{aligned}
\end{equation}
In the above, the inner limit of the amplitude function $\Phi$ from equation \eqref{eq:ampsol1} has been taken by defining
\begin{equation}\label{eq:strangeP}
\mathcal{P}(\sigma)=\int_{0}^{\sigma}\bigg[ \frac{a \i \phi_1^{\prime}(t)}{[\phi_0^{\prime}(t)]^3}-\frac{3}{2(t-a \i)}\bigg] \mathrm{d}t,
\end{equation}
such that $\mathcal{P}(\sigma)=O(1)$ as $\sigma \to a \i$.
Matching \eqref{eq:outerlimit} with \eqref{eq:innerlimit}, and substituting for $\alpha=1/2$ from \eqref{eq:constfop}, determines the constant, $\Lambda$, as
\begin{equation}\label{eq:constfopapp}
\Lambda = -\frac{a \i \e^{-\mathcal{P}(a \i)}}{3 \sqrt{3}} \lim_{n \to \infty} \bigg(\frac{\hat{\phi}_n}{\Gamma(n+\alpha+1)}\bigg).
\end{equation}

% \newpage

\end{document}